\documentclass[runningheads]{llncs}
\usepackage[T1]{fontenc}
\usepackage{graphicx,verbatim}
\usepackage{amsmath}
\usepackage{amssymb}
\usepackage{multirow}
\usepackage{colortbl}
\usepackage{booktabs}
\usepackage[table]{xcolor}
\usepackage{marvosym}

\begin{document}
\title{Automatic LV Localization and Short-Axis Plane Estimation from Arbitrary CMR Slice}
\titlerunning{LV Localization and SAX Plane Estimation from Arbitrary Slice}
%
\author{Yi Yu\inst{1} \and
Yixuan Liu\inst{1} \and
Ziyu Zhang\inst{2} \and
Parker Martin\inst{1} \and
Zhenyu Bu\inst{1} \and \\
Yuchi Han\inst{1} \and
Yuan Xue\inst{1}\textsuperscript{(\Letter)}}
   

\authorrunning{Y. Yu et al.}
%
\institute{
The Ohio State University, Columbus, USA\\
\email{Yuan.Xue@osumc.edu} \\
\and
Nanjing University, Nanjing, China 
}


  
\maketitle              
\begin{abstract}

Accurate estimation of left ventricular (LV) orientation is essential for cardiac magnetic resonance (CMR) imaging and downstream analysis. Existing methods typically formulate orientation recognition as discrete view classification or rely on multi-slice geometric intersection, limiting their ability to model continuous 3D orientation and generalize across arbitrary slices. This work introduces a novel paradigm: Joint LV localization and 3D orientation estimation from a single CMR slice. To investigate this setting, representative orientation-aware detection frameworks are adapted to the CMR domain, and their limitations are analyzed. Upon that, we propose the Polar-Coupled Circular (PCC) embedding that provides a continuous and unambiguous orientation representation to address the limitations. Meanwhile, a scalable benchmark is constructed through automatic slice sampling from volumetric CMR segmentation datasets. Extensive experiments on four datasets demonstrate strong performance, achieving an average mIoU of 86.18\% and an average angle deviation of 3.39$^\circ$. This study establishes a new task setting for single-slice LV orientation modeling and provides a geometry-consistent framework for spatially informed CMR analysis. Code is available at \url{https://github.com/yuyi1005/cmr-3d-ood}.


\end{abstract}

\section{Introduction}

Cardiovascular magnetic resonance (CMR) plays a central role in quantitative cardiac assessment, providing high-resolution anatomical and functional information across multiple imaging planes \cite{yu2026uncertainty}. In particular, evaluation of the left ventricle (LV) is fundamental for measuring cardiac function and diagnosing disease. Over the past decade, LV segmentation and quantification have been extensively studied and achieved substantial progress. These tasks highly rely on proper cardiac orientation, especially short-axis (SAX) orientation. Consequently, accurate modeling of \textbf{3D SAX orientation} is increasingly important for spatially aware analysis \cite{bagci2011orientation,kellman2011lvlocalization}, supporting applications such as cross-view registration, cardiac modeling, and automated plane prescription \cite{bottcher2025fully}.

\begin{figure}[t]
\includegraphics[width=\textwidth]{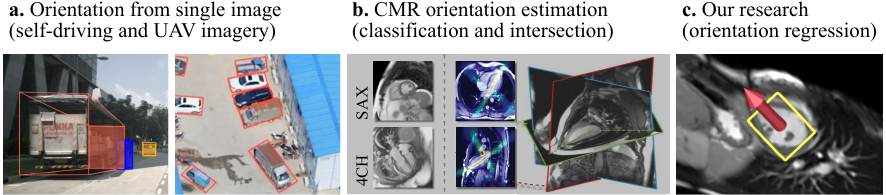}
\caption{Orientation recognition from a single image has achieved remarkable success in autonomous driving. However, in the CMR domain, orientation estimation still largely relies on traditional view classification or plane-intersection strategies.} \label{fig:related}
\end{figure}

Existing research on SAX orientation recognition, however, largely follows two paradigms: \textbf{1)} Some studies formulate it as a \textbf{view classification} problem, assigning images to predefined categories such as SAX, four-chamber (4CH), or two-chamber (2CH) views \cite{aung2019classification,vergani2021classification,chauhan2022view}. While effective for coarse categorization, they are inherently limited to discrete classes and cannot model continuous 3D orientation. \textbf{2)} Another line of work focuses on automated plane prescription, which estimates the cardiac orientation during acquisition \cite{glessgen2025autovsmanual}. They estimate the target plane through \textbf{geometric intersection} across multiple slices (see Fig.~\ref{fig:related}b) \cite{lu2011view,blansit2019deep,edalati2022implementation,wei2021view}. Such approaches require multi-view aggregation.

Recent advances in autonomous driving have demonstrated that directly regressing 3D orientation from 2D observations is both feasible and highly effective. For instance, FCOS3D \cite{wang2021fcos3d} can reliably localize vehicles and infer their 3D orientation from monocular images (see Fig.~\ref{fig:related}a). In CMR, despite the absence of perspective projection, different slice orientations likewise exhibit distinct anatomical patterns.
This raises a natural question: Can 3D oriented object detection be extended to CMR? In this work, we explore this new direction and demonstrate that accurate LV localization and 3D orientation estimation from a single slice are achievable with strong performance (see Fig.~\ref{fig:related}c).

Our contributions are summarized as:
\textbf{1) Task setting and benchmark.} We introduce a new problem setting: LV localization and 3D orientation estimation from a single CMR slice, and develop a data curation pipeline to construct the benchmark from CMR segmentation datasets.
\textbf{2) Analysis of potential solutions.} We adapt four representative orientation-aware detection frameworks to the CMR setting, establishing strong baselines. Their behaviors are systematically analyzed to identify limitations in 3D SAX orientation estimation.
\textbf{3) Methodological contribution.} To address the limitations of existing methods, we propose Polar-Coupled Circular (PCC) embedding that enforces unit-energy consistency, improving stability and accuracy in 3D orientation estimation. 

\begin{figure}[t]
\includegraphics[width=\textwidth]{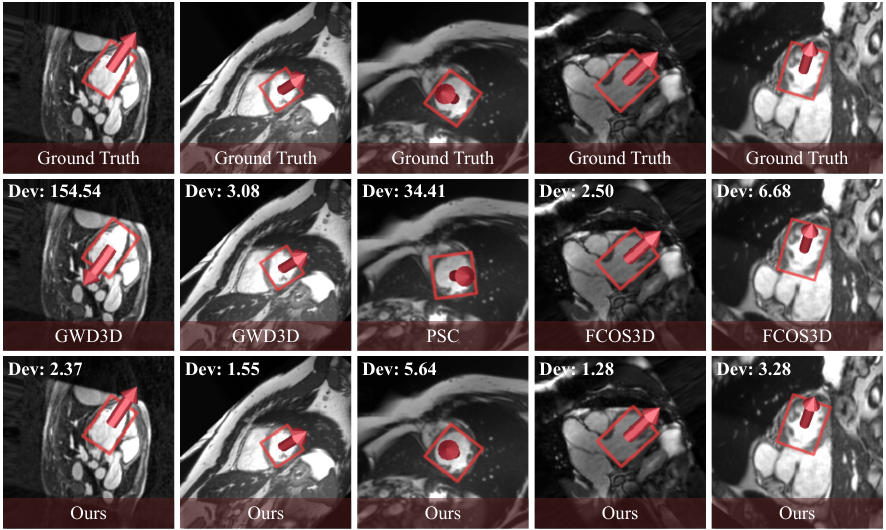}
\caption{Visualization of our work. Our model localizes the left ventricle with a rotated bounding box and estimates its 3D orientation (red arrows) using a single, arbitrary 2D slice. The second row displays some compared methods repurposed to our task.} \label{fig:vis}
\end{figure}

\section{Analysis of Applicable Methods}\label{sec:related}

3D oriented object detection is an active research area, particularly in autonomous driving, where models estimate object orientation and spatial extent directly from images \cite{ma2024object}. Representative methods include: 
\textbf{1) L1-based.}
Early methods directly regress 3D box parameters, including rotation angles, using L1 or smooth L1 losses. For example, FCOS3D~\cite{wang2021fcos3d} predicts orientation explicitly in parameter space, treating angle estimation as standard regression.
\textbf{2) IoU-based.}
Intersection over Union (IoU) optimizes geometric overlap instead of independent parameters. RIoU3D~\cite{zheng2020rotation} introduces differentiable rotated IoU to supervise alignment directly in 3D geometric space.
\textbf{3) Gaussian-based.}
Gaussian-based methods reformulate oriented boxes as 2D/3D Gaussian distributions. GWD~\cite{yang2023detecting} measures similarity using Gaussian Wasserstein Distance, implicitly encoding orientation in the covariance matrix.
\textbf{4) Encoding-based.}
They encode oriented boxes into and regress encoded embeddings. PSC~\cite{yu2023psc,yu2024boundary} proposes boundary-free encoding to improve rotation stability during optimization.

\noindent\textbf{Limitation analysis.}
The applicability of existing orientation-aware detection methods to CMR has not been previously studied. We therefore adapt representative approaches to our task. For methods originally designed for 2D rotation (e.g., PSC), a z-channel is added to extend them to 3D. As shown in Fig.~\ref{fig:vis}, several limitations arise:
\textbf{1) Ambiguity in IoU and Gaussian.}
These methods are invariant to 180° rotation, making opposite directions indistinguishable. As a result, they cannot represent the full semi-spherical orientation space, leading to reversed predictions in certain cases.
\textbf{2) Boundary issue in L1-based regression.}
Due to angular periodicity, $180^\circ$ and $-180^\circ$ are equivalent but yield a large L1 loss. This artificial discontinuity causes unstable optimization and degraded accuracy near boundaries.
\textbf{3) Instability of PSC embedding.}
Embedding-based methods such as PSC alleviate 2D boundary problem but are not inherently designed for 3D orientation. Extending them with an additional z-channel leads to varying embedding magnitudes across different directions and instability near polar regions (e.g., $z \to 1$), resulting in unreliable predictions.

Overall, existing formulations work well in LV localization, but exhibit representation ambiguity, boundary discontinuity, or instability when extended to continuous 3D SAX orientation regression in CMR.

\section{Methodology}

By analyzing the failure cases of existing methods (see Sec.~\ref{sec:related}), an ideal orientation model should satisfy the following properties:
\textbf{1) Full coverage.} Its orientation spans the entire semi-sphere.
\textbf{2) Uniform magnitude.} The embedding $\mathbf{n}$ satisfies $\|\mathbf{n}\|_2 = 1$ for all orientations.
\textbf{3) Continuity.} No discontinuities across angular boundaries.
\textbf{4) Uniqueness.} One-to-one correspondence between embedding and physical orientation, avoiding ambiguity.

\begin{figure}[t]
\includegraphics[width=\textwidth]{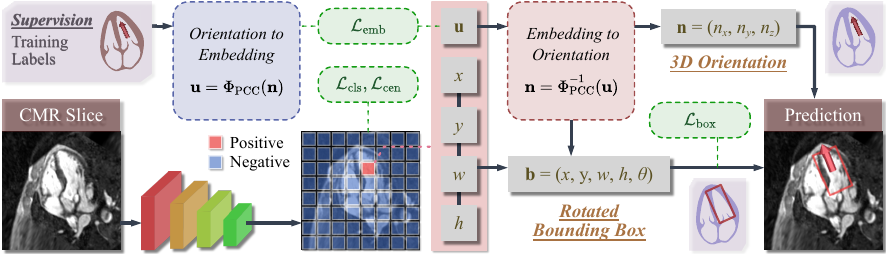}
\caption{The framework of our proposed model.} \label{fig:framework}
\end{figure}

In this section, we present our modified encoding-based method with all these features (see Fig. \ref{fig:framework}). 
The network adopts a ResNet-50 \cite{he2016deep} with FPN \cite{lin2017feature} backbone, following the FCOS framework \cite{tian2019fcos}. In addition to the standard bounding box parameters $(x, y, w, h)$, we introduce a Polar-Coupled Circular (PCC) embedding $\mathbf{u}$ for orientation prediction.
During training, the embedding is computed from the ground-truth 3D normal vector as
$\mathbf{u} = \Phi_{\mathrm{PCC}}(\mathbf{n})$,
as detailed in Sec.~\ref{sec:method-pcc}. The network is supervised to regress $\mathbf{u}$.
During inference, the predicted embedding is decoded to recover the orientation
$\mathbf{n} = \Phi_{\mathrm{PCC}}^{-1}(\mathbf{u})$,
as described in Sec.~\ref{sec:method-dec}. The in-plane angle is then computed as
$\theta = \arctan2(n_y, n_x)$,
and combined with $(x, y, w, h)$ to form the rotated bounding box $(x, y, w, h, \theta)$.

\subsection{Polar-Coupled Circular (PCC) Embeddings}\label{sec:method-pcc}

To obtain a continuous and geometrically consistent representation of 3D orientation, we introduce an PCC embedding, denoted as
$\mathbf{u} = \Phi_{\mathrm{PCC}}(\mathbf{n})$.

Given a unit normal vector $\mathbf{n} = (n_x, n_y, n_z)$, we encode it using evenly spaced circular harmonic bases. Let $N$ be the encoding length. For $N\ge4$, we compute:
\begin{equation}
p_n = \sqrt{\frac{2}{N-1}\left(1 - n_z^2\right)} \cos\!\left(\theta + \frac{2\pi n}{N-1}\right),
\quad n = 0, \dots, N-2
\label{equ:enc}
\end{equation}
where $\theta = \arctan2(n_y, n_x) $ and the last term $p_{N-1} = n_z$. The final embedding is constructed as:
\begin{equation}
\mathbf{u} = \left[p_0, \cdots, p_{N-1} \right]^\top
\end{equation}

For the special case of $N=3$, the embedding is directly defined as $\mathbf{u}=(n_x,n_y,n_z)$, corresponding to the Cartesian representation of the normal vector. The scaling factor in Eq. (\ref{equ:enc}) couples the azimuthal components with the polar coordinate, making the embedding satisfy the unit-energy constraint $\|\mathbf{u}\|_2^2 = 1$, ensuring that $\mathbf{u}$ lies on the unit hypersphere. This polar coupling suppresses azimuthal instability as $z \rightarrow 1$, where orientation about the polar axis becomes ambiguous, and provides a smooth and geometrically consistent representation for stable 3D orientation regression.

\subsection{Recover Orientation from PCC Embeddings}\label{sec:method-dec}

Given a PCC embedding $\mathbf{u} = [p_0, \cdots, p_{N-1}]^\top$, the reverse process $\mathbf{n} = \Phi_{\mathrm{PCC}}^{-1}(\mathbf{u})$ recovers the 3D orientation vector $\mathbf{n} = (n_x, n_y, n_z)$ by projecting the circular harmonic components onto the first-order cosine and sine bases. Let $N$ denote the encoding length. For $N\ge4$, we compute the orthogonal projection:
\begin{equation}
(n_x,n_y)
=
\sqrt{\frac{2}{N-1}}
\sum_{n=0}^{N-2}
p_n
\left(
\cos\frac{2\pi n}{N-1},
-\sin\frac{2\pi n}{N-1}
\right)
\end{equation}

The polar component is directly preserved as $n_z = p_{N-1}$. The resulting vector $\mathbf{n} = (n_x, n_y, n_z)$ recovers the Cartesian representation of the 3D orientation. 

\subsection{Loss Functions}

Based on FCOS \cite{tian2019fcos}, our method inherited a classification loss (Focal loss \cite{lin2017focal}), a box regression loss (IoU loss), and a center-ness loss (Binary Cross-Entropy loss). When PCC is incorporated, it involves an embedding loss. In the following, we introduce its calculation. First, the ground truth normal is encoded as:
\begin{equation}
\mathbf{u}_\text{GT} = \Phi_{\mathrm{PCC}}(\mathbf{n}_\text{GT})
\end{equation}
where $\mathbf{u}_\text{GT}$ is the ground truth embeddings.

Afterward, the embedding loss can be calculated with $L_1$ loss:
\begin{equation}
\mathcal{L}_\text{emb}=L_1 \left ( \mathbf{u}_\text{pred}, \mathbf{u}_\text{GT} \right )
\end{equation}
where $\mathbf{u}_\text{pred}$ is the output embeddings of the network.

Finally, with the weighted losses $w_\text{cls}\mathcal{L}_\text{cls}$, $w_\text{box}\mathcal{L}_\text{box}$, and $w_\text{cen}\mathcal{L}_\text{cen}$ defined by backbone FCOS detector, the overall loss for training can be expressed as:
\begin{equation}
\mathcal{L}=w_\text{cls}\mathcal{L}_\text{cls} + w_\text{box}\mathcal{L}_\text{box} + w_\text{cen}\mathcal{L}_\text{cen} + w_\text{emb}\mathcal{L}_\text{emb}
\label{eq:lossall}
\end{equation}
where $w_\text{emb}$ is set to 2 according to our ablation study (see Sec. \ref{sec:exp}). 

\subsection{Scalable Training Data Construction}\label{sec:method-curation}

\begin{figure}[t]
\includegraphics[width=\textwidth]{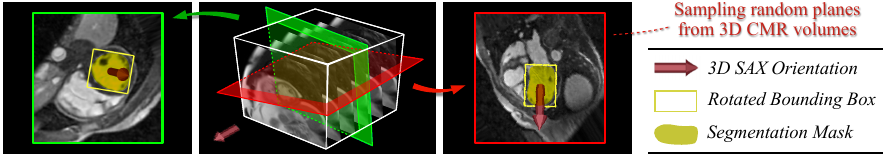}
\caption{Illustration of the data generation pipeline. Readily available 3D segmentation datasets are repurposed for our task setting, LV localization and oriented detection.} \label{fig:datacuration}
\end{figure}

For a newly defined task, the first and most essential contribution is the benchmark establishment. However, manually annotating 3D orientation from 2D images is extremely challenging. To address this limitation, we develop a data processing pipeline that automatically constructs a benchmark from volumetric CMR segmentation datasets, which are more widely available.

Given a 3D CMR volume with LV segmentation mask and the SAX direction (which is known for SAX stacks and manually annotated otherwise), we construct training samples through stochastic slicing (see Fig.~\ref{fig:datacuration}). Specifically, we first randomly select a point $P$ inside the LV mask. A normal vector $\mathbf{v}$, uniformly sampled from the unit sphere, is then used to define a plane $\Pi$ passing through $P$. Since $P$ lies within the LV mask, plane $\Pi$ is guaranteed to intersect the LV, ensuring visible cardiac structures in the resulting slice. The volume is resampled onto plane $\Pi$ to obtain a $256 \times 256$ 2D CMR image, with the in-plane spatial resolution uniformly sampled between 1\,mm and 2\,mm. The original 3D SAX direction is then transformed into the coordinate system of this plane, yielding 3D orientation labels. For localization supervision, the minimum rotated bounding box enclosing LV mask is computed. The rotation of the box is aligned with the projection of the SAX direction onto plane $\Pi$.

By repeating this procedure, numerous randomly oriented 2D slices can be sampled from a single 3D volume. Given the availability of large-scale public segmentation datasets, this strategy is naturally scalable. In our experiments, 20 slices are sampled from each mask-annotated 3D CMR volume.

\begin{table*}[!tb]
\fontsize{8.5pt}{10pt}\selectfont
\setlength{\tabcolsep}{5.2mm}
\caption{Comparison between our method and existing approaches adapted to our task. Orientation accuracy is measured by angle deviation (in degrees, reported as mean $\pm$ std), localization accuracy by mIoU, and detection performance by AP$_{50}$ and AP$_{75}$, evaluated on the ACDC, M\&Ms-2, In-House, and V2S-Real datasets.}
\label{tab:exp_main}
\centering
\begin{tabular}{l|c|c|c|c}
\toprule
{\textbf{Methods}} & \textbf{Angle Dev.} & \textbf{mIoU} & \textbf{AP}$_{50}$ & \textbf{AP}$_{75}$ \\
\hline
\rowcolor{gray!20} \multicolumn{5}{l}{$\blacktriangledown$ \textit{ACDC Dataset}} \\ \hline
RIoU3D (2020) \cite{zheng2020rotation} & 57.84 $\pm$ 46.51 & 82.94 & 99.27 & 74.17 \\
FCOS3D (2021) \cite{wang2021fcos3d} & 4.87 $\pm$ 3.33 & 83.57 & 99.10 & 77.63\\
GWD3D (2023) \cite{yang2023detecting} & 58.73 $\pm$ 45.19 & 83.09 & 99.23 & 74.38 \\
PSC (2024) \cite{yu2024boundary} & 4.49 $\pm$ 3.50 & 83.15 & 99.10 & 78.55 \\
\textbf{Ours} & \textbf{3.73 $\pm$ 2.82} & \textbf{83.96} & \textbf{99.11} & \textbf{80.47} \\ \hline
\rowcolor{gray!20} \multicolumn{5}{l}{$\blacktriangledown$ \textit{M\&Ms-2 Dataset}} \\ \hline
RIoU3D (2020) \cite{zheng2020rotation} & 51.33 $\pm$ 40.77 & 83.66 & 99.89 & 78.18 \\
FCOS3D (2021) \cite{wang2021fcos3d} & 4.94 $\pm$ 3.84 & 84.12 & 99.78 & 81.44 \\
GWD3D (2023) \cite{yang2023detecting} & 52.26 $\pm$ 39.20 & 83.59 & 99.92 & 76.65 \\
PSC (2024) \cite{yu2024boundary} & 4.67 $\pm$ 3.10 & 84.01 & 99.84 & 82.29 \\
\textbf{Ours} & \textbf{4.04 $\pm$ 3.01} & \textbf{84.08} & \textbf{99.86} & \textbf{82.91} \\ \hline
\rowcolor{gray!20} \multicolumn{5}{l}{$\blacktriangledown$ \textit{In-House Dataset}} \\ \hline
RIoU3D (2020) \cite{zheng2020rotation} & 57.36 $\pm$ 46.94 & 86.73 & 99.98 & 88.67 \\
FCOS3D (2021) \cite{wang2021fcos3d} & 4.94 $\pm$ 3.84 & 84.22 & 99.78 & 81.44 \\
GWD3D (2023) \cite{yang2023detecting} & 58.09 $\pm$ 45.29 & 86.62 & 99.99 & 87.53 \\
PSC (2024) \cite{yu2024boundary} & 3.47 $\pm$ 2.89 & 87.03 & 99.94 & 91.02 \\
\textbf{Ours} & \textbf{3.05 $\pm$ 2.42} & \textbf{87.15} & \textbf{99.99} & \textbf{91.51} \\ \hline
\rowcolor{gray!20} \multicolumn{5}{l}{$\blacktriangledown$ \textit{V2S-Real Dataset}} \\ \hline
RIoU3D (2020) \cite{zheng2020rotation} & 62.21 $\pm$ 50.47 & 87.87 & 99.65 & 91.95 \\
FCOS3D (2021) \cite{wang2021fcos3d} & 4.65 $\pm$ 3.57 & 89.07 & 99.62 & 94.71 \\
GWD3D (2023) \cite{yang2023detecting} & 63.72 $\pm$ 47.10 & 88.06 & 99.75 & 90.59 \\
PSC (2024) \cite{yu2024boundary} & 3.52 $\pm$ 2.61 & 89.14 & 99.70 & 96.42 \\
\textbf{Ours} & \textbf{2.75 $\pm$ 2.07} & \textbf{89.54} & \textbf{99.70} & \textbf{96.95} \\
\bottomrule
\end{tabular}
\end{table*}

\section{Experiments}\label{sec:exp}

Experiments are carried out on NVIDIA A100 GPUs using PyTorch 2.4.0 \cite{paszke2019pytorch} and the rotation detection toolkit MMRotate 1.0.0 \cite{zhou2022mmrotate}. All the experiments follow the same hyper-parameters (learning rate, batch size, optimizer, etc.). Mean Intersection over Union (mIoU) and Average Precision (AP) are used for localization, and angle deviation for 3D orientation. All models adopt a ResNet-50 backbone \cite{he2016deep} for fair comparisons. Training is performed with AdamW \cite{loshchilov2018decoupled} for 36 epochs, with an initial learning rate of $1e{-4}$, 500-iteration warm-up, and $\times$0.1 decay at epochs 24 and 33. Random rotation is used for data augmentation. Training and testing data are curated using our pipeline in Sec.~\ref{sec:method-curation}.

\noindent\textbf{Datasets. }
We combine all training sets to train a unified model for release, and evaluate it separately on the test sets of the four datasets:
\textbf{1) ACDC \cite{bernard2018acdc}.} It contains 150 SAX CMR CINE scans (100 training / 50 testing). End-diastole (ED) and end-systole (ES) frames are annotated with masks.
\textbf{2) M\&Ms-2 \cite{campello2021mnms2,martin2023mnms2}.} This multi-center dataset includes 360 SAX CMR CINE scans with ED/ES annotations, with 160 scans held out for testing.
\textbf{3) In-House.} It contains 188 CINE scans (95 for testing) annotated on all 30 frames, ensuring the trained model generalizes across arbitrary cardiac phases.
\textbf{4) V2S-Real.} It contains 2,000 in-house isotropic 3D frames from 10 volunteers across 10 cardiac cycles, with the first frame manually annotated and the remaining frames propagated using \cite{zhang2025atlas}. The isotropic resolution avoids resampling degradation and enables realistic arbitrary slicing for evaluation.

\begin{table*}[!tb]
\fontsize{8.5pt}{10pt}\selectfont
\setlength{\tabcolsep}{5mm}
\setlength{\aboverulesep}{0.4ex}
\setlength{\belowrulesep}{0.4ex}
\caption{Ablation studies with the length of encoding and the weight of embedding loss on V2S-Real dataset.}
\label{tab:exp_ablation}
\centering
\begin{tabular}{c|c|c||c|c|c}
\toprule
{\textbf{\textit{N}}} & \textbf{Angle Dev.} & \textbf{AP}$_{75}$ & \textbf{\textit{w}}$_\text{emb}$ & \textbf{Angle Dev.} & \textbf{AP}$_{75}$ \\ \hline
3 & 2.99 $\pm$ 2.15 & 96.13 & 0.5 & 3.27 $\pm$ 2.79 & 95.89 \\
4 & \textbf{2.75 $\pm$ 2.07} & \textbf{96.95} & 1.0 & 2.91 $\pm$ 2.14 & 96.26 \\
7 & 3.09 $\pm$ 2.60 & 94.68 & 2.0 & \textbf{2.75 $\pm$ 2.07} & \textbf{96.95} \\
13 & 3.16 $\pm$ 2.67 & 93.63 & 5.0 & 3.15 $\pm$ 3.32 & 89.07 \\
\bottomrule
\end{tabular}
\end{table*}

\begin{figure}[t]
\includegraphics[width=\textwidth]{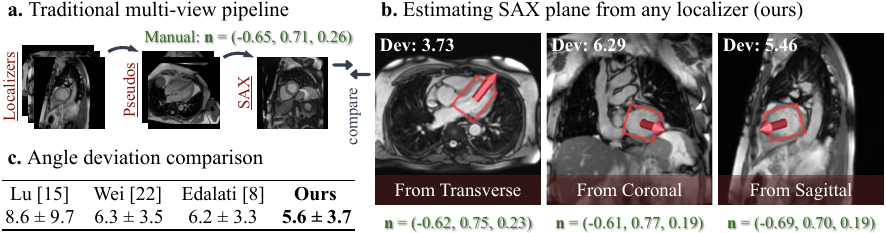}
\caption{Comparison on the real-world SAX plane prescription task. We compare our predicted SAX normal with manually prescribed planes, demonstrating performance on par with existing automatic multi-view plane prescription methods \cite{lu2011view,wei2021view,edalati2022implementation}.} \label{fig:exp}
\end{figure}

\noindent\textbf{Main results. }
Results are shown in Tab.~\ref{tab:exp_main}, with visualization shown in Fig.~\ref{fig:vis}. IoU- and Gaussian-based methods achieve competitive localization accuracy, but exhibit large angle deviations due to their inherent 180$^\circ$ rotational ambiguity. L1-based FCOS3D covers the full angular range, yet suffers from boundary discontinuity. PSC alleviates the in-plane boundary issue, but remains unstable as $z \rightarrow 1$. In contrast, our method resolves these limitations and achieves the best overall performance (averaged over four datasets): mIoU of 86.18\% (0.35\% higher than PSC), AP$_{75}$ of 87.96\% (0.89\% higher), and angle deviation of 3.39$^\circ$ (0.65$^\circ$ lower). Owing to its continuous and stable representation over the full orientation range, it also yields lower standard deviation.

\noindent\textbf{Ablation studies.} The embedding length $N=4$ in PCC and the weight of the embedding loss $w_\text{emb}=2$ are determined according to Tab.~\ref{tab:exp_ablation}.

\noindent\textbf{Comparison with multi-view.}
We evaluate four volunteers in a real-world clinical prescription setting. Following the clinical workflow, localizers are first acquired, and pseudo 2CH/4CH views are subsequently captured for manual multi-view estimation of the SAX normal. The reference normal is derived from the SAX DICOM metadata of the clinical process, providing reliable ground truth for comparison with our predictions (see Fig.~\ref{fig:exp}a--b). The summarized angle deviations in Fig.~\ref{fig:exp}c demonstrate the effectiveness of the proposed approach.

\section{Conclusion}

We present a novel paradigm for single-slice LV localization and 3D orientation estimation in CMR, offering a simpler alternative to view classification and geometric intersection approaches. To support this task, we develop a scalable data curation pipeline to build a dedicated benchmark. We adapt four representative orientation-aware detection frameworks and analyze their limitations when applied to SAX orientation estimation in CMR. To address these issues, we introduce a Polar-Coupled Circular (PCC) embedding mechanism that provides a continuous and unambiguous orientation representation. Extensive experiments on four datasets demonstrate strong performance, achieving an average mIoU of 86.18\%, an angle deviation of 3.39$^\circ$, and an AP$_{75}$ of 87.96\%.

    

\begin{credits}
\subsubsection{\ackname} 
This study was supported in part by The Ohio State University Clinical and Translational Science Institute (CTSI) and the National Center for Advancing Translational Sciences of the National Institutes of Health (NIH) under Grant UM1TR004548, and by NIH Grant R01 HL148103.

\subsubsection{\discintname}
The authors have no competing interests to declare that
are relevant to the content of this article.
\end{credits}

%
%
%
\bibliographystyle{splncs04}
\bibliography{mybibliography}
%




\end{document}